\documentclass[aps,twocolumn,letterpaper,superscriptaddress]{revtex4-1}

\usepackage{amsmath}
\usepackage{amssymb}
\usepackage{amsfonts}
\usepackage{bm}
\usepackage{mathrsfs}
\usepackage{tensor}
\usepackage{dsfont}

\usepackage[usenames,dvipsnames]{xcolor}
\usepackage[hyperindex,pdftex, breaklinks,colorlinks = true,linkcolor = blue,urlcolor=blue,citecolor=blue]{hyperref}

\usepackage{physics}
\usepackage[section]{placeins}

\begin{document}
	
	\title{Polarization and orbital magnetization in Chern insulators: A microscopic perspective}
	
	\author{Perry T. Mahon}
	\email{perry.mahon@austin.utexas.edu}
	\affiliation{Department of Physics, University of Toronto, Toronto, Ontario M5S 1A7, Canada}
	\affiliation{Department of Physics, University of Texas at Austin, Austin, Texas 78712, USA}
	
	\author{Jason G. Kattan}
	\email{jkattan@physics.utoronto.ca}
	\affiliation{Department of Physics, University of Toronto, Toronto, Ontario M5S 1A7, Canada}
	
	\author{J. E. Sipe}
	\email{sipe@physics.utoronto.ca}
	\affiliation{Department of Physics, University of Toronto, Toronto, Ontario M5S 1A7, Canada}
	
	\date{\today }
	
	\begin{abstract}
		We derive macroscopic expressions for the polarization and orbital magnetization of a Chern insulator in its zero-temperature ground state using a previously developed formalism for treating microscopic polarization and magnetization fields in extended media. In the limit of a topologically trivial insulator, our results reduce to those of the ``modern theories of polarization and magnetization.'' In a Chern insulator, however, we find a generically nonvanishing microscopic free current density, the macroscopic average of which vanishes. Moreover, the expression that we obtain for the polarization is qualitatively similar to that of the ``modern theory,'' while the expressions for the orbital magnetization fundamentally differ; the manner in which they differ elucidates the distinct philosophies of these theoretical frameworks. 
		\iffalse
		In a Chern insulator, the polarization takes a familiar form, although the ambiguity in its definition is no longer quantized. The form of the magnetization is also modified, also admitting a general ambiguity in its value. Such gauge dependency is not problematic within our theory, as we do not view such intermediate quantities, from which the expectation values of the electronic change and current density operators can be extracted, as being directly physically accessible.
		\fi
	\end{abstract}
	
	\maketitle

	\section{Introduction}
	At the start of the last century, the electrical conduction of metals was understood as arising from the response of ``free charges,'' while in insulators with solely ``bound charges'' only a change in the polarization could arise were an electric field applied, and no persistent current would result \cite{Lorentz}. In the successor quantum treatment developed in the 1930s \cite{Ashcroft}, metals were associated with partly occupied energy bands, where an electric field could induce current flow through a redistribution of Bloch electrons in the Brillouin zone, while insulators were associated with occupied bands that are energetically separated from unoccupied bands by a band gap, and no such redistribution could occur. The link with the earlier picture of insulators was only established near the end of the last century, with the ``modern theories of polarization and magnetization'' \cite{Vanderbilt1993, Resta1994}. Here the electronic polarization of an insulator is associated with the dipole moment of Wannier functions, some localized near each lattice site, constructed from superpositions of the occupied (cell-periodic) energy eigenfunctions, and the change in this polarization as an electric field is applied can be calculated.
	
	Yet it is currently appreciated that this narrative is too naive. In a so-called Chern insulator, for example, the zero-temperature electronic ground state is such that there is a set of $N$ occupied energy bands separated from the unoccupied bands by an energy gap, and in that sense it is an ``insulating phase." But even in the absence of an applied magnetic field, such insulators break time-reversal symmetry internally, and a current can be induced by an applied electric field, albeit one perpendicular to the electric field. In a two-dimensional Chern insulator, the (quantized) Hall conductivity is \cite{TKNN}
	\begin{align*}
		& \sigma^{xy}=\frac{e^{2}}{2\pi\hbar} C_{\mathcal{V}},
	\end{align*}
	where $e=-|e|$ is the electron charge, and $C_{\mathcal{V}}\in\mathbb{Z}$ is identified as the (first) Chern number, which is given by
	\begin{align*}
		& C_{\mathcal{V}}=\frac{1}{2\pi}\sum_{n=1}^{N}\int_{\text{BZ}}d\boldsymbol{k}\left(\frac{\partial\xi_{nn}^{y}(\boldsymbol{k})}{\partial k_{x}} - \frac{\partial\xi_{nn}^{x}(\boldsymbol{k})}{\partial k_{y}}\right).
	\end{align*}
	Here $d\boldsymbol{k}=dk_{x}dk_{y}$, and the integrand involves derivatives of the diagonal components of the non-Abelian Berry connection $\xi^a(\boldsymbol{k})$ associated with the cell-periodic parts of the Bloch energy eigenfunctions. The relevance of quantities like $C_{\mathcal{V}}$ caused a paradigm shift in condensed-matter theory, and it has become appreciated that it is not simply the spectral properties of a crystal's band structure that are important but also its topological properties \footnote{See, \textit{e.g.}, Simon \cite{Simon1983}}. To capture these properties, the electronic Hilbert space must be decomposed into a family of subspaces $\{\mathcal{H}_{\bm{k}}\}_{\bm{k}\in\text{BZ}}$ parameterized by the (first) Brillouin zone (BZ) and spanned pointwise by a collection of cell-periodic Bloch states $\{\ket{n\bm{k}}\}_{n \in I}$ (corresponding to the cell-periodic parts
	$u_{n\bm{k}}(\bm{x})\equiv\bra{\bm{x}}\ket{n\bm{k}}$ of the Bloch energy eigenfunctions), where $I \subseteq \mathbb{N}$ indexes some isolated set of bands, all of which are assembled into a mathematical structure called a \textit{Hilbert bundle} over the BZ \cite{Panati2017}. Indeed, $C_{\mathcal{V}}$ is a topological invariant of one such Hilbert bundle (where the index set $I$ runs only over the occupied bands of an insulator), and if this quantity is nonvanishing, then that Hilbert bundle is nontrivial \footnote{See, \textit{e.g.}, Chapter 11 of Nakahara \cite{NakaharaBook}}; crystalline insulators supporting such a nonvanishing $C_{\mathcal{V}}$ are thus deemed ``topologically nontrivial.'' Another consequence of this nontriviality is that exponentially localized Wannier functions (ELWFs) \textit{cannot} be constructed from superpositions of the occupied Bloch energy eigenfunctions alone \cite{Brouder,Monaco2017}. Therefore the response of a Chern insulator to an electric field cannot be addressed in the same way as that of a topologically trivial insulator.
	
	And so how does one make sense of concepts like ``polarization'' and ``magnetization'' for Chern insulators? The ``modern theories'' use adiabatically induced currents to define these quantities, where it is implicitly assumed that free charge and current densities vanish \footnote{See, \textit{e.g.}, Resta \cite{Resta2010}}. These approaches focus on directly identifying the macroscopic polarization and magnetization, and extensions to treat topologically nontrivial insulators typically involve thermodynamic arguments \cite{Niu2007}; justification of the resulting expressions is argued through numerical comparisons with the polarization and orbital magnetization of finite-sized samples \cite{Resta2006,Vanderbilt2009}. In contrast, we have introduced an approach based on defining \textit{microscopic} polarization and magnetization fields in generic crystalline solids, which are constructed using ELWFs, and understand the macroscopic fields as arising from the spatial averages of those microscopic fields \cite{Mahon2019}. Here, polarization and magnetization fields serve as intermediary quantities that aid calculation and provide physical insight, but in general only the appropriate combinations that lead to the charge and current densities have direct physical significance. This approach has been implemented to study the effect of applied electromagnetic fields that can vary arbitrarily in space and time \cite{Mahon2020,Mahon2020a}, and recently we have used it to study the metallic systems that arise from $p$-doping trivial insulators \cite{Mahon2021}. Notably, the expressions for the polarization and magnetization
	in topologically trivial insulators derived within this framework coincide with those of the ``modern theories.''
	
	In this paper we implement this microscopic approach to treat the polarization
	and magnetization of a Chern insulator in its ground state. To construct the ELWFs upon which we base our microscopic description, we employ not only the occupied energy eigenfunctions, but include sufficiently many unoccupied energy eigenfunctions so that such a construction is permitted. We obtain different expressions for the macroscopic ground-state polarization and magnetization than
	those obtained from the ``modern theories,'' and we also find that
	a microscopic free current density exists; the spatial average of this quantity is found to vanish, consistent with assumptions therein. But we argue that the nonvanishing of this microscopic quantity, which is sensitive to the topology of the valence bands, serves as an indicator that a particular material is not a trivial insulator. With the ground state so characterized, the stage is set for a full consideration of quantities induced within a Chern insulator by arbitrary electromagnetic fields.
	
	\section{Polarization and Magnetization}
	\subsection{Microscopic quantities}\label{Sec2A}
	To begin, consider a three-dimensional bulk insulator in the independent particle approximation, occupying its zero-temperature electronic ground state $\ket{\text{gs}}$, for which we identify two distinct sets of cell-periodic Bloch states. The first set consists of all of the states $\ket{n\bm{k}}$ with corresponding energies $E_{n\boldsymbol{k}}$ less than the Fermi energy $E_{F}$; suppose that there are $N$ such \textit{valence} bands. These energy bands will generally intersect each other, and we are interested in a scenario in which the Hilbert bundle constructed from the collection $\{\ket{n\bm{k}}\}_{n=1}^N$ for each $\bm{k}\in\text{BZ}$, called the \textit{valence bundle} $\mathcal{V}$ \footnote{Here and henceforth we refer to a given Hilbert bundle over the Brillouin zone by its total space.}, is topologically nontrivial. In two dimensions, this occurs if and only if the (first) Chern number $C_{\mathcal{V}}$ is nonzero \cite{Panati2017}, while in three dimensions there are a triple of such Chern numbers $C_{\mathcal{V}}^i$ ($i=1,2,3$), at least one of which must be nonzero \footnote{The $i^{\text{th}}$ Chern number $C_{\mathcal{V}}^i$ is defined to be the integral of the (first) Chern class over the submanifold of $\text{BZ}$ obtained by fixing the $i^{\text{th}}$ coordinate $k^i$, which is diffeomorphic to a $2$-torus.}. The second set consists of all of the states $\ket{n\bm{k}}$ associated with sufficiently many conduction bands, which similarly span the fibers of a Hilbert bundle called the \textit{conduction bundle} $\mathcal{C}$, such that the \textit{Bloch bundle} $\mathcal{B} = \mathcal{V} \oplus \mathcal{C}$ \footnote{This means that the Bloch bundle is isomorphic as a Hilbert bundle to the Whitney sum $\mathcal{V} \oplus \mathcal{C}$ of the valence and conduction bundles $\mathcal{V}$ and $\mathcal{C}$.}, constructed from the states associated with all of these bands (indexed by a set $J \subseteq \mathbb{N}$), is topologically trivial \cite{PanatiPisante}. In two dimensions, this means that the Chern number $C_{\mathcal{C}}$ of the conduction bundle is opposite to that of the valence bundle; that is, the second set is defined such that $C_{\mathcal{V}}+C_{\mathcal{C}}=0$. In three dimensions we require that $C_{\mathcal{V}}^i + C_{\mathcal{C}}^i = 0$ for all $i=1,2,3$. Notably, it has been shown that such a set always exists, although in general it must consist of all of the conduction bands (in which case $J = \mathbb{N}$) \cite{Freed}. Hilbert bundles that are topologically trivial admit a globally defined orthonormal basis (or \textit{frame}) for their fibers that is smooth and (in this case) periodic, a necessary condition for the construction of a set of ELWFs \footnote{In addition to each ELWF needing to be spatially localized, the set of ELWFs must be orthonormal and span a Hilbert space that is isomorphic to that spanned by the set of energy eigenfunctions used in their construction.}. In general, for each $\boldsymbol{k}\in\text{BZ}$ the components of such a ``Wannier frame" $(\ket{\alpha\bm{k}})_{\alpha \in J}$ can be obtained by a unitary transformation of the components of the (only locally defined) ``Bloch frame" $(\ket{n\bm{k}})_{n \in J}$,
	\begin{align}
		\ket{\alpha\boldsymbol{k}}=\sum_{n}U_{n\alpha}(\boldsymbol{k})\ket{n\boldsymbol{k}},
		\label{pseudoBloch}
	\end{align}
	where the sum is over all $n \in J$. The corresponding cell-periodic functions $u_{\alpha\boldsymbol{k}}(\boldsymbol{x})\equiv\braket{\boldsymbol{x}}{\alpha\boldsymbol{k}}$ can be transformed into a set of ELWFs via \cite{Brouder,Vanderbilt2011,Marzari2012,PanatiPisante,Troyer2016}
	\begin{align}
		W_{\alpha\boldsymbol{R}}(\boldsymbol{x})=\sqrt{\Omega_{\text{uc}}}\int_{\text{BZ}}\frac{d\boldsymbol{k}}{(2\pi)^3}e^{i\boldsymbol{k}\boldsymbol{\cdot}(\boldsymbol{x}-\boldsymbol{R})}u_{\alpha\boldsymbol{k}}(\boldsymbol{x}),
		\label{WFfull}
	\end{align}
	where $\alpha$ is a ``type'' index, $\Omega_{\text{uc}}$ denotes the volume of the unit cell $\Omega$ over which the cell-periodic functions are normalized, and $\boldsymbol{R}$ is an element of a Bravais lattice $\Gamma$ that characterizes the underlying crystal structure of the material.
	
	We work in the frozen-ion approximation, taking the ion cores to be fixed, and neglect the electronic spin degree of freedom; therefore in the Heisenberg picture the electron field operators encode the only dynamical degrees of freedom of the crystal. Their dynamical evolution, governed by the Heisenberg equation
	\begin{equation}
		i \hbar \frac{\partial \hat{\psi}(\bm{x},t)}{\partial t} = \big[\hat{\psi}(\bm{x},t), \hat{\mathsf{H}}_0\big],
	\end{equation}
	is generated by the Schrödinger operator
	\begin{equation}
		\hat{\mathsf{H}}_0 = \int d\bm{x}\, \hat{\psi}^{\dagger}(\bm{x},t) \mathcal{H}_0(\bm{x}) \hat{\psi}(\bm{x},t),\label{H0}
	\end{equation}
	involving the $\Gamma$-periodic differential operator
	\begin{equation}
		\mathcal{H}_0(\bm{x}) = \frac{1}{2m} \big(\bm{\mathfrak{p}}(\bm{x})\big)^2 + \mathrm{V}_{\Gamma}(\bm{x}).
	\end{equation}
	Here $\mathrm{V}_{\Gamma}(\bm{x})$ is the electrostatic potential associated with the ion cores that characterizes the crystal structure and satisfies $\mathrm{V}_{\Gamma}(\bm{x}+\bm{R}) = \mathrm{V}_{\Gamma}(\bm{x})$ for all $\bm{R} \in \Gamma$, and
	\begin{equation}
		\bm{\mathfrak{p}}(\bm{x}) = \frac{\hbar}{i} \bm{\nabla} - \frac{e}{c} \bm{A}_{\text{static}}(\bm{x}),
		\label{frakturp}
	\end{equation}
	where $\bm{A}_{\text{static}}(\bm{x})$ is an ``internal," cell-periodic vector potential that generally breaks time-reversal symmetry in the unperturbed crystal.
	
	We introduce fermionic creation and annihilation operators $\hat{a}_{n\bm{k}}^{\dagger}$ and $\hat{a}_{n\bm{k}}$ (obeying the usual anticommutation relations) for which $\ket{\psi_{n\boldsymbol{k}}}\equiv \hat{a}^\dagger_{n\boldsymbol{k}}\ket{\text{vac}}$, where $\hat{\mathsf{H}}_0\ket{\psi_{n\boldsymbol{k}}}=E_{n\boldsymbol{k}}\ket{\psi_{n\boldsymbol{k}}}$. The Bloch energy eigenfunctions are defined by $\psi_{n\bm{k}}(\bm{x}) \equiv \bra{\bm{x}}\ket{\psi_{n\bm{k}}}$ and have related cell-periodic functions $u_{n\bm{k}}(\bm{x})\equiv\bra{\bm{x}}\ket{n\bm{k}}$ through Bloch's theorem:
	\begin{equation}
		\psi_{n\bm{k}}(\bm{x}) = \frac{1}{(2\pi)^{3/2}} e^{i\bm{k}\cdot\bm{x}}u_{n\bm{k}}(\bm{x}).\label{blochfunctions}
	\end{equation}
	Under the dynamical evolution generated by $\hat{\mathsf{H}}_0$, 
	\begin{equation}
		\hat{a}_{n\boldsymbol{k}}(t)=e^{-iE_{n\boldsymbol{k}}t/\hbar}\hat{a}_{n\boldsymbol{k}}.
		\label{fieldoperatordynamics}
	\end{equation}
	
	We also introduce fermionic creation and annihilation operators $\hat{a}_{\beta\bm{R}'}^{\dagger}$ and $\hat{a}_{\beta\bm{R}'}$ (also obeying the usual anticommutation relations), for which $\ket{\beta\bm{R}'} \equiv \hat{a}_{\beta\bm{R}'}^{\dagger}\ket{\text{vac}}$, with the ELWFs $W_{\beta\bm{R}'}(\bm{x}) \equiv \bra{\bm{x}}\ket{\beta\bm{R}'}$ being the associated coordinate functions. Through (\ref{pseudoBloch},\ref{WFfull}) and (\ref{blochfunctions}) we find the relationship between the fermionic operators generating ELWFs $\hat{a}^\dagger_{\beta\boldsymbol{R}'}$ and those generating the Bloch energy eigenfunctions $\hat{a}^\dagger_{n\boldsymbol{k}}$, namely,
	\begin{equation}
		\hat{a}_{\beta\bm{R}'}^{\dagger}(t) = \sqrt{\frac{\Omega_{\text{uc}}}{(2\pi)^3}} \int_{\text{BZ}} d\bm{k}\, e^{-i\bm{k}\cdot\bm{R}'} \sum_n U_{\beta n}(\bm{k}) \hat{a}_{n\bm{k}}^{\dagger}(t).
	\end{equation}
	From the perspective of our microscopic theory, the primary quantity capturing the difference between a ``trivial'' insulator and a Chern insulator is the (electronic) single-particle density matrix $\eta_{\alpha\boldsymbol{R}'';\beta\boldsymbol{R}'}(t)$ \cite{Mahon2019}. In the case of an unperturbed insulator occupying its zero-temperature ground state $\ket{\text{gs}}$ considered here, the general definition for this quantity simplifies such that $\eta_{\alpha\boldsymbol{R}'';\beta\boldsymbol{R}'}(t)\rightarrow\eta_{\alpha\beta}(\boldsymbol{R}''-\boldsymbol{R}',t)$, and moreover
	\begin{align}
		&\eta_{\alpha\beta}(\boldsymbol{R}''-\boldsymbol{R}',t) \nonumber\\
		&=\bra{\text{\text{gs}}}\hat{a}^\dagger_{\beta\boldsymbol{R}'}(t)\hat{a}_{\alpha\boldsymbol{R}''}(t)\ket{\text{\text{gs}}}\nonumber\\
		&=\frac{\Omega_{\text{uc}}}{(2\pi)^3}\int_{\text{BZ}}d\boldsymbol{k}\,e^{i\boldsymbol{k}\boldsymbol{\cdot}(\boldsymbol{R}''-\boldsymbol{R}')}\sum_{n}f_nU^\dagger_{\alpha n}(\boldsymbol{k})U_{n\beta}(\boldsymbol{k}),
		\label{EDM}
	\end{align}
	which is independent of time; $f_n=\Theta(E_F-E_{n\boldsymbol{k}})$ is the occupation factor of the Bloch energy eigenvector $\ket{\psi_{n\boldsymbol{k}}}$ and is independent of $\boldsymbol{k}$ in the case of an insulator.
	
	The limit of a topologically trivial insulator occurs when $C_{\mathcal{V}}^i = 0$ for all $i=1,2,3$, in which case the Bloch energy eigenfunctions associated with each set of isolated bands can be mapped to a subset of the complete set of ELWFs. In particular, we can require that the states $\ket{\alpha\boldsymbol{k}}$ be associated with cell-periodic functions obtained from superpositions of either the occupied \emph{or} the unoccupied cell-periodic energy eigenfunctions exclusively; this renders the $U_{n\alpha}(\boldsymbol{k})$ block diagonal with ``upper block'' of size $N\times N$ \footnote{This means that the states $\ket{\alpha\bm{k}}$ live exclusively in the fibers of either the valence \textit{or} conduction bundle for all $\alpha$}. We can then introduce an analogous occupation factor $f_{\alpha}$ associated with $\ket{\alpha\boldsymbol{k}}$; we set the $f_{\alpha}$ associated with $\ket{\alpha\boldsymbol{k}}$ to be equal to the $f_n$ associated with the $\ket{n\boldsymbol{k}}$ used in its construction. It then follows that, in this limit,
	\begin{equation*}
		\eta^{\text{(trivial)}}_{\alpha\boldsymbol{R}'';\beta\boldsymbol{R}'}=f_{\alpha}\delta_{\alpha\beta}\delta_{\boldsymbol{R}''\boldsymbol{R}'},
	\end{equation*}
	as expected \cite{Mahon2019,Mahon2020}.
	
	In earlier work \cite{Mahon2019,Mahon2020} we showed that, under the frozen-ion approximation, the total microscopic charge and current densities of an extended system with electronic degrees of freedom minimally coupled to an arbitrary electromagnetic field can be written as
	\begin{subequations}
		\begin{align}
			\rho(\boldsymbol{x},t)&=-\boldsymbol{\nabla}\boldsymbol{\cdot}\boldsymbol{p}(\boldsymbol{x},t)+\rho_{\text{F}}(\boldsymbol{x},t),\\
			\boldsymbol{j}(\boldsymbol{x},t)&=\frac{\partial\boldsymbol{p}(\boldsymbol{x},t)}{\partial t}+c\boldsymbol{\nabla}\times\boldsymbol{m}(\boldsymbol{x},t)+\boldsymbol{j}_{\text{F}}(\boldsymbol{x},t),
		\end{align}
		\label{eq:micro}%
	\end{subequations}
	where $\bm{p}(\bm{x},t)$ and $\bm{m}(\bm{x},t)$ are microscopic polarization and magnetization fields, and $\rho_{\text{F}}(\bm{x},t)$ and $\bm{j}_{\text{F}}(\bm{x},t)$ are free charge and current densities. Under that approximation, $\rho(\bm{x},t)=\langle\hat{\rho}(\boldsymbol{x},t)\rangle+\rho^{\text{ion}}(\boldsymbol{x})$ and $\boldsymbol{j}(\bm{x},t)=\langle\hat{\boldsymbol{j}}(\boldsymbol{x},t)\rangle$, with $\hat{\rho}(\boldsymbol{x},t)$ and $\hat{\boldsymbol{j}}(\boldsymbol{x},t)$ the electronic charge and current density operators arising from the conserved Noether current. To identify the quantities on the right-hand side of (\ref{eq:micro}), we first decomposed $\rho(\boldsymbol{x},t)$ and $\boldsymbol{j}(\boldsymbol{x},t)$ as a sum of spatially localized contributions, one associated with each lattice site $\bm{R}\in\Gamma$. The static charge density $\rho^{\text{ion}}(\bm{x})$ describing the distribution of ion cores naturally decomposes as
	\begin{equation}
		\rho^{\text{ion}}(\bm{x}) = \sum_{\boldsymbol{R}}\rho_{\boldsymbol{R}}^{\text{ion}}(\boldsymbol{x}),
	\end{equation}
	where, for example, taking the ions to be pointlike,
	\begin{equation*}
		\rho_{\bm{R}}^{\text{ion}}(\bm{x}) = \sum_N q_N \delta(\bm{x} - \bm{R} - \bm{d}_N).
	\end{equation*}
	The sum is over ion cores in the unit cell, with $q_N$ denoting the charge of the $N^{\text{th}}$ ion core located at $\bm{R} + \bm{d}_N$. To identify localized portions of the (expectation values of the) electronic charge and current densities, we utilize a complete set of ELWFs with respect to which they are decomposed as a sum of ``site" electronic charge and current densities,
	\begin{subequations}
		\begin{align}
			\langle\hat{\rho}(\boldsymbol{x},t)\rangle &=\sum_{\boldsymbol{R}}\rho^{\text{el}}_{\boldsymbol{R}}(\boldsymbol{x},t), \\
			\langle\hat{\boldsymbol{j}}(\boldsymbol{x},t)\rangle &=\sum_{\boldsymbol{R}}\boldsymbol{j}_{\boldsymbol{R}}(\boldsymbol{x},t).
		\end{align}
	\end{subequations}
	And, indeed, the ``site'' quantities $\rho^{\text{el}}_{\boldsymbol{R}}(\boldsymbol{x},t)$ and $\boldsymbol{j}_{\boldsymbol{R}}(\boldsymbol{x},t)$ are nonvanishing only for $\bm{x}$ ``near'' $\boldsymbol{R}$; they can be written in terms of the single-particle density matrix as
	\begin{align}
		\rho_{\bm{R}}^{\text{el}}(\bm{x},t) = \sum_{\alpha\beta\bm{R}'\bm{R}''} \rho_{\beta\bm{R}';\alpha\bm{R}''}(\bm{x},\bm{R};t)\, \eta_{\alpha\bm{R}'';\beta\bm{R}'}(t), \\
		\bm{j}_{\bm{R}}(\bm{x},t) = \sum_{\alpha\beta\bm{R}'\bm{R}''} \bm{j}_{\beta\bm{R}';\alpha\bm{R}''}(\bm{x},\bm{R};t)\, \eta_{\alpha\bm{R}'';\beta\bm{R}'}(t),
	\end{align}
	involving the previously introduced \cite{Mahon2019} ``generalized electronic site-quantity matrix elements" $\rho_{\beta\bm{R}';\alpha\bm{R}''}(\bm{x},\bm{R};t)$ and $\bm{j}_{\beta\bm{R}';\alpha\bm{R}''}(\bm{x},\bm{R};t)$. From the ``site'' charge and current densities, site polarization and magnetization fields can be defined in a manner similar to that of atomic and molecular physics. In particular, the site polarization field is taken to be
	\begin{equation}
		\bm{p}_{\bm{R}}(\bm{x},t) = \int d\bm{y}\, \bm{s}(\bm{x};\bm{y},\bm{R}) \Big(\rho_{\bm{R}}^{\text{el}}(\bm{y},t) + \rho_{\bm{R}}^{\text{ion}}(\bm{y})\Big),
	\end{equation}
	where we have introduced a ``relator" \cite{Mahon2019}, defined by the distributional expression
	\begin{equation}
		\bm{s}(\bm{x};\bm{y},\bm{R}) = \int_{C(\bm{y},\bm{R})} d\bm{z}\, \delta(\bm{x} - \bm{z}),
	\end{equation}
	with $C(\bm{y},\bm{R})$ denoting an arbitrary continuously differentiable curve that begins at $\bm{R}$ and ends at $\bm{y}$. The total microscopic polarization field is then defined as
	\begin{equation}
		\bm{p}(\bm{x},t) \equiv \sum_{\bm{R}} \bm{p}_{\bm{R}}(\bm{x},t).
		\label{microP}
	\end{equation}
	Meanwhile, the site magnetization field is expressed as a sum of two contributions: The familiar, ``atomiclike'' site magnetization field is
	\begin{equation}
		\bar{m}_{\bm{R}}^i(\bm{x},t) = \frac{1}{c}\int d\bm{y}\, \alpha^{ib}(\bm{x};\bm{y},\bm{R}) j_{\bm{R}}^b(\bm{y},t),
	\end{equation}
	where we have defined another ``relator" \cite{Mahon2019},
	\begin{equation}
		\alpha^{ij}(\bm{x};\bm{y},\bm{R}) = \epsilon^{imn} \int_{C(\bm{y},\bm{R})} dz^m\, \frac{\partial z^n}{\partial y^j} \delta(\bm{x}-\bm{z});
	\end{equation}
	here and below, superscript indices indicate Cartesian components, and repeated Cartesian components are summed over. But unlike in an isolated atom or a molecular crystal, there is generally a local nonconservation of electronic charge and current densities associated with each site,
	\begin{equation}
		\frac{\partial \rho_{\bm{R}}^{\text{el}}(\bm{x},t)}{\partial t} + \bm{\nabla}\cdot\bm{j}_{\bm{R}}(\bm{x},t) \neq 0,
	\end{equation}
	which arises because, although the total electronic charge-current density is conserved by construction, ELWFs associated with different $\bm{R}\in\Gamma$ may have common support at a given point $\bm{x}$. This results in an additional ``itinerant" contribution to the total magnetization field,
	\begin{equation}
		\tilde{m}_{\bm{R}}^i(\bm{x},t) = \frac{1}{c} \int d\bm{y}\, \alpha^{ib}(\bm{x};\bm{y},\bm{R}) \tilde{j}_{\bm{R}}^{b}(\bm{y},t),
	\end{equation}
	and it has been shown that $\tilde{\bm{j}}_{\bm{R}}(\bm{x},t)$ can also be written in terms of the single-particle density matrix
	\begin{equation}
		\tilde{\bm{j}}_{\bm{R}}(\bm{x},t) = \sum_{\alpha\beta\bm{R}'\bm{R}''} \tilde{\bm{j}}_{\beta\bm{R}';\alpha\bm{R}''}(\bm{x},\bm{R};t)\, \eta_{\alpha\bm{R}'';\beta\bm{R}'}(t),
		\label{itinerantcurrentsitequant}
	\end{equation}
	where the ``generalized site-quantity matrix element" $\tilde{\bm{j}}_{\beta\bm{R}';\alpha\bm{R}''}(\bm{x},\bm{R};t)$ was also defined previously \cite{Mahon2019}. The total microscopic magnetization field is then defined as
	\begin{equation}
		\bm{m}(\bm{x},t) \equiv \sum_{\bm{R}} \Big(\bar{\bm{m}}_{\bm{R}}(\bm{x},t) + \tilde{\bm{m}}_{\bm{R}}(\bm{x},t)\Big).
		\label{microM}
	\end{equation}
	\\
	From the site polarization and magnetization fields, site dipole moments can be extracted \cite{Mahon2020} via
	\begin{subequations}
		\begin{align}
			\bm{\mu}_{\bm{R}}(t) &= \int d\bm{x}\, \bm{p}_{\bm{R}}(\bm{x},t), \\
			\bm{\nu}_{\bm{R}}(t) &= \int d\bm{x}\, \bm{m}_{\bm{R}}(\bm{x},t).
		\end{align}
		\label{multipole}%
	\end{subequations}
	
	\subsection{Macroscopic quantities}
	Here we study an insulator occupying its zero-temperature ground state and do not consider the effect of an applied electromagnetic field. Thus all quantities appearing in (\ref{eq:micro}-\ref{multipole}) become independent of time. Moreover, the electric and magnetic dipole moments $\boldsymbol{\mu}_{\boldsymbol{R}}$ and $\boldsymbol{\nu}_{\boldsymbol{R}}$ associated with each lattice site are physically equivalent; $\boldsymbol{\mu}_{\boldsymbol{R}}=\boldsymbol{\mu}_{\boldsymbol{R}'}$ and $\boldsymbol{\nu}_{\boldsymbol{R}}=\boldsymbol{\nu}_{\boldsymbol{R}'}$ for any pair of lattice sites $\boldsymbol{R},\boldsymbol{R}' \in\Gamma$. Consequently, the macroscopic polarization and magnetization fields are uniform,
	\begin{equation}
		\boldsymbol{P}=\frac{\boldsymbol{\mu}_{\boldsymbol{R}}}{\Omega_{\text{uc}}} \; \; \; \text{and} \; \; \;
		\boldsymbol{M}=\frac{\boldsymbol{\nu}_{\boldsymbol{R}}}{\Omega_{\text{uc}}}, \label{PM}
	\end{equation}
	where the electric and magnetic dipole moments are given explicitly in terms of the single-particle density matrix by
	\begin{widetext}
		\begin{subequations}
			\begin{align}
				\mu^i_{\boldsymbol{R}} &= \sum_{\alpha\beta\boldsymbol{R}'\boldsymbol{R}''}\bigg[\int d\bm{y}\,\big(y^i-R^i\big)\rho_{\beta\boldsymbol{R}';\alpha\boldsymbol{R}''}(\boldsymbol{y},\boldsymbol{R})\bigg]\eta_{\alpha\beta}(\boldsymbol{R}''-\boldsymbol{R}') + \big(\boldsymbol{\mu}_{\boldsymbol{R}}^{\text{ion}}\big)^i, \label{dipoleMoments_mu}\\
				\nu^i_{\boldsymbol{R}} &= \sum_{\alpha\beta\boldsymbol{R}'\boldsymbol{R}''}\bigg[\frac{\epsilon^{iab}}{2c}\int d\bm{y}\,\big(y^a-R^a\big)\Big(j_{\beta\boldsymbol{R}';\alpha\boldsymbol{R}''}^{b}(\boldsymbol{y},\boldsymbol{R}) + \tilde{j}_{\beta\boldsymbol{R}';\alpha\boldsymbol{R}''}^{b}(\boldsymbol{y},\boldsymbol{R})\Big)\bigg]\eta_{\alpha\beta}(\boldsymbol{R}''-\boldsymbol{R}'), \label{dipoleMoments_nu}
			\end{align}
		\end{subequations}
	\end{widetext}
	where $\boldsymbol{\mu}_{\boldsymbol{R}}^{\text{ion}}$ is the dipole moment associated with $\rho_{\boldsymbol{R}}^{\text{ion}}(\boldsymbol{x})$, and the ``generalized site-quantity matrix elements" in their time-independent form are given in Appendix \ref{appendixA}.
	
	By implementing (\ref{EDM}) in (\ref{dipoleMoments_mu}), we obtain the expression for the macroscopic polarization of a Chern insulator, which is found to be (see Appendix \ref{appendixB})
	\begin{equation}
		P^i=e\sum_{n}f_n\int_{\text{BZ}}\frac{d\boldsymbol{k}}{(2\pi)^3}\Big(\xi^i_{nn}(\bm{k})+\mathcal{W}^i_{nn}(\bm{k})\Big)+\frac{\big(\boldsymbol{\mu}_{\boldsymbol{R}}^{\text{ion}}\big)^i}{\Omega_{\text{uc}}},
		\label{mu}
	\end{equation}
	where the components of the non-Abelian Berry connection in the locally defined ``Bloch frame" $(\ket{n\bm{k}})_n$ are
	\begin{equation}
		\xi_{mn}^a(\bm{k}) = \frac{i}{\Omega_{\text{uc}}} \int_{\Omega} d\bm{x}\, u_{m\bm{k}}^*(\bm{x})\frac{\partial u_{n\bm{k}}(\bm{x})}{\partial k^a},
		\label{eq:Bloch_connection}
	\end{equation}
	and we have defined the Hermitian matrix \cite{VanderbiltBook} populated by elements
	\begin{equation}
		\mathcal{W}^a_{mn}(\boldsymbol{k})\equiv i\sum\limits_{\alpha}\big(\partial_aU_{m\alpha}(\boldsymbol{k})\big)U^\dagger_{\alpha n}(\boldsymbol{k}). \label{W}
	\end{equation}
	The expression (\ref{mu}) is formally similar to that of a trivial insulator {\cite{Resta1994}}. As described in Sec.~\ref{Sec2A}, in a trivial insulator $U_{n\alpha}(\boldsymbol{k})$ can be chosen to be of block-diagonal form, and under such circumstances the term in (\ref{mu}) involving $\mathcal{W}^i_{nn}(\bm{k})$ has been shown to generally evaluate to an element of a discrete set \cite{Resta1994}; that is, there is a ``quantum of indeterminacy'' associated with $\bm{P}$. However, for a Chern insulator the transformation (\ref{pseudoBloch}) must involve states associated with both the valence and conduction bands. Thus $U_{n\alpha}(\boldsymbol{k})$ cannot be chosen to be of that block-diagonal form, and the ambiguity associated with (\ref{mu}) is not generally discrete, at least not following from the usual argument of Resta \cite{Resta1994}. Nevertheless, the macroscopic polarization (\ref{mu}) maintains the physically sensible feature that shifting the origin of all ELWFs by any $\boldsymbol{R}_{\text{s}} \in \Gamma$ leads to a shift in $\boldsymbol{P}$ by an additive constant proportional to $\bm{R}_s$. Explicitly, mapping $\ket{\alpha\boldsymbol{R}}\rightarrow\ket{\alpha\boldsymbol{R}+\boldsymbol{R}_{\text{s}}}$, or equivalently, $U_{n\alpha}(\bm{k})\rightarrow e^{-i\boldsymbol{k}\cdot\boldsymbol{R}_{\text{s}}}U_{n\alpha}(\bm{k})$ and thus $\mathcal{W}^a_{mn}(\bm{k}) \rightarrow\mathcal{W}^a_{mn}(\bm{k})+\delta_{mn}R^a_{\text{s}}$, leads to a shift
	\begin{equation}
		\boldsymbol{P}\rightarrow\boldsymbol{P}+eN_{\text{el}}\boldsymbol{R}_{\text{s}},
	\end{equation}
	where $N_{\text{el}}$ is the number of electrons per unit volume.
	
	Unlike the polarization (\ref{mu}), the expression that we obtain for the macroscopic orbital magnetization is qualitatively different from that of the ``modern theory'' \cite{Resta2006,Niu2007}. From (\ref{dipoleMoments_nu}), there are two distinct contributions thereto; we will call the first of these the \textit{atomiclike} contribution (indicated by a bar accent) and the second the \textit{itinerant} contribution (indicated by a tilde accent). For a Chern insulator, the atomiclike contribution is
	\begin{widetext}
		\begin{equation}
			\bar{M}^i=-\frac{e}{2\hbar c}\epsilon^{iab}\text{Im}\sum_{nm}f_n\int_{\text{BZ}}\frac{d\boldsymbol{k}}{(2\pi)^3}\Big(E_{m\boldsymbol{k}}\xi^a_{nm}\xi^b_{mn}+\big(E_{m\boldsymbol{k}}-E_{n\boldsymbol{k}}\big)\mathcal{W}^a_{nm}\xi^b_{mn}-E_{n\boldsymbol{k}}\mathcal{W}^a_{nm}\mathcal{W}^b_{mn}\Big),
			\label{nuAtomic}
		\end{equation}
		which reduces to the usual atomic term in the limit of a trivial insulator, while the itinerant contribution is
		\begin{equation}
			\tilde{M}^i=-\frac{e}{2\hbar c}\epsilon^{iab}\text{Im}\sum_{n}f_n\int_{\text{BZ}}\frac{d\boldsymbol{k}}{(2\pi)^3}\Big(-iE_{n\boldsymbol{k}}\partial_a\xi^b_{nn}+\sum_{m}\big(E_{n\boldsymbol{k}}-E_{m\boldsymbol{k}}\big)\mathcal{W}^a_{nm}\xi^b_{mn}-E_{m\boldsymbol{k}}\mathcal{W}^a_{nm}\mathcal{W}^b_{mn}\Big),
			\label{nuItinerant}
		\end{equation}
		which also reduces to the usual itinerant term in that limit \cite{Resta2005,Mahon2020}. Combining (\ref{nuAtomic}) and (\ref{nuItinerant}), we find
		\begin{equation}
			M^i=\frac{e}{2\hbar c}\sum_{n}f_n\int_{\text{BZ}}\frac{d\boldsymbol{k}}{(2\pi)^3}\Big(E_{n\boldsymbol{k}}\epsilon^{iab}\partial_a\xi^b_{nn}-\sum_{m}E_{m\boldsymbol{k}}\text{Im}\big[\epsilon^{iab}\xi^a_{nm}\xi^b_{mn}\big]\Big)+\frac{e}{2\hbar c}\sum_{nm}(f_n-f_m)\int_{\text{BZ}}\frac{d\boldsymbol{k}}{(2\pi)^3}E_{n\boldsymbol{k}}\text{Im}\big[\epsilon^{iab}\mathcal{W}^a_{nm}\mathcal{W}^b_{mn}\big].
			\label{nu}
		\end{equation}
	\end{widetext}
	The first integral in (\ref{nu}) constitutes the usual expression for the orbital magnetization of a trivial insulator, while the second is sensitive to the global topology of the underlying valence bundle. 
	
	To illustrate this, consider first a generic trivial insulator. As described above, the $U_{n\alpha}(\bm{k})$ through which we construct ELWFs can be chosen to be block diagonal, so that $\mathcal{W}_{mn}^a(\bm{k})\neq 0$ only if $f_m=f_n$. The second term of (\ref{nu}) then vanishes, and the usual expression is obtained \cite{Resta2005,Mahon2020}. Consider instead the simplest instance of a Chern insulator, in which the energy bands are isolated from one another; even in this case, ELWFs can only be constructed from the energy eigenfunctions associated with the full Bloch bundle. Thus the $U_{n\alpha}(\bm{k})$ will necessarily have at least four off-diagonal entries, two associated with $n,\alpha\in\{1,\ldots,N\}$ and two associated with $n,\alpha\in\{N+1,\ldots\}$. Hence it is clear from (\ref{W}) that in general $\mathcal{W}_{mn}^a(\bm{k}) \neq 0$ for $f_m \neq f_n$, in which case the second term of (\ref{nu}) need not vanish.
	
	In fact, this second integral is in disagreement with the generalization of the ``modern theory" to Chern insulators \cite{Resta2006,Niu2007}. There the orbital magnetization of a Chern insulator is obtained by adding to the analogous expression for a trivial insulator a term involving the product of the Berry curvature and a chemical potential; in Resta \textit{et al.} \cite{Resta2006} it is argued that this form is invariant under transformations between local Bloch frames for the valence bundle $\mathcal{V}$ and manifestly exhibits invariance of $\boldsymbol{M}$ under a shift of the energy zero. We too anticipate (\ref{mu}) and (\ref{nu}) to be unaffected by shifts of the energy zero, $E_{n\boldsymbol{k}}\rightarrow E_{n\boldsymbol{k}}+\varepsilon$ and $E_{F}\rightarrow E_{F}+\varepsilon$. The polarization (\ref{mu}) is trivially unaffected by such a shift, while the magnetization (\ref{nu}) becomes
	\begin{widetext}
		\begin{equation}
			M^i\rightarrow M^i+\varepsilon\frac{e}{\hbar c}\sum_{n}f_n\int_{\text{BZ}}\frac{d\boldsymbol{k}}{(2\pi)^3}\epsilon^{iab}\Big(\partial_a \xi^b_{nn}(\bm{k}) + \partial_a\mathcal{W}^b_{nn}(\bm{k})\Big).
			\label{energyShift}
		\end{equation}
	\end{widetext}
	It is not obvious, \textit{a priori}, that the second term in (\ref{energyShift}) vanishes, nor would we expect it to vanish for an arbitrary multiband unitary transformation $U(\bm{k})$. However, as shown in Appendix \ref{appendixC}, the integral does vanish for those $\mathcal{W}_{mn}^a(\bm{k})$ related to unitary transformations $U_{n\alpha}(\boldsymbol{k})$, whose action on the local Bloch frame $(\ket{n\boldsymbol{k}})_{n \in J}$ results in a smooth, globally defined Wannier frame $(\ket{\alpha\boldsymbol{k}})_{\alpha \in J}$ for the Bloch bundle.
	
	The vanishing of the integral in (\ref{energyShift}) provides us with an interpretation of the matrix $\mathcal{W}_{mn}^a(\bm{k})$, an object that is ubiquitous in the formalism developed above. The first term in the parentheses of (\ref{energyShift}) is directly proportional to the Chern number $C^{i}_{\mathcal{V}}$, which encodes information about how this Hilbert bundle ``twists'' as elements therein are parallel transported between neighboring fibers using the non-Abelian Berry connection $\xi_{mn}^a(\bm{k})$. The existence of a set of ELWFs constructed from elements of the Bloch bundle implies that the integrand itself vanishes \textit{locally}, meaning that any such ``twisting'' is locally ``unwound'' by the matrix $\mathcal{W}_{mn}^a(\bm{k})$ in such a way that the globally defined Wannier frame $(\ket{\alpha\bm{k}})_{\alpha \in J}$ is smooth and periodic everywhere. Degeneracies within the valence bands act as obstructions to extending this local cancellation across the entire Brillouin zone \cite{Monaco2017}. However, when integrated over the Brillouin zone, the sum total of the contributions coming from these obstructions cancels among the various terms in the integrand \cite{Kaufmann}, and therefore leads to the vanishing of the integral in (\ref{energyShift}). In this sense there exists here a delicate cancellation between terms associated with the \textit{local topology} coming from band crossings, and the \textit{global topology} associated with the ``twisting'' of the valence bundle itself.
	
	\subsection{Comparison with existing literature}
	\label{Sec:2C}
	
	As detailed above, in constructing macroscopic quantities to describe the physics of bulk Chern insulators, our formalism is to be distinguished from the approach taken in the ``modern theories" (or rather their generalizations to topologically nontrivial insulators). In the latter, following a solution of the spectral problem associated with the Hamiltonian (\ref{H0}) through Bloch's theorem, one uses the cell-periodic parts of the occupied Bloch states $\{\ket{n\bm{k}}\}_{n=1}^N$, or the associated projectors
	\begin{equation}
		P_{\mathcal{V}}(\bm{k}) = \sum_{n} f_n \ketbra{n\bm{k}},
	\end{equation}
	to calculate the Berry curvature and the (first) Chern numbers $C^{i}_{\mathcal{V}}$ characterizing the valence bundle $\mathcal{V}$. If any such integer is nonzero, then the standard conclusion is to abandon the construction of ELWFs and instead use thermodynamic arguments to extend, for example, the definition of the macroscopic (bulk) orbital magnetization in topologically trivial insulators to their nontrivial counterparts, usually by introduction of a chemical potential lying somewhere in the band gap. While this is perfectly reasonable in finite-size systems, for bulk systems with no boundary, the location of the chemical potential within the band gap is essentially arbitrary. Conversely, in our formalism, after including sufficiently many unoccupied conduction bands so that the total Chern numbers $C^{i}_{\mathcal{B}}=C^{i}_{\mathcal{V}} + C^{i}_{\mathcal{C}}$ characterizing the valence and conduction bundles taken together (the Bloch bundle) vanish, we are then able to construct ELWFs using a unitary transformation of the form (\ref{pseudoBloch}) and thereby build an inherently microscopic description of the crystal, obtaining macroscopic quantities like the polarization and magnetization through spatial averaging. Of course, we have replaced the arbitrariness in placement of a chemical potential with a sort of ``gauge freedom" in the choice of Wannier frame $(\ket{\alpha\bm{k}})_{\alpha \in J}$ (or associated ELWFs) that is used to define the site quantities of Sec.~\ref{Sec2A}. This type of ``gauge freedom" is outside of the scope of the description of Chern insulators given by the ``modern theories." The gauge freedom considered there consists only of transformations between local Bloch frames of the valence bundle $\mathcal{V}$, which (away from degeneracies) take the form $\ket{n\bm{k}}\rightarrow e^{i\lambda_{n}(\bm{k})}\ket{n\bm{k}}$ for some smooth function $\lambda_n(\bm{k})$. Under such transformations, they find their expression for the magnetization to be invariant, as is ours. However, unlike in the ``modern theories," we would argue that our ``gauge freedom" is an advantage, since it can be used to minimize the spread of the ELWFs --- for example, using the Marzari-Vanderbilt functional \cite{Marzari2012} --- allowing one to improve the accuracy of various approximations associated with, for example, the construction of tight-binding models using our choice of ELWFs. 
	
	Of course, in focusing on bulk crystals for which it is the charge and current densities that are generally physically accessible, which are related to the microscopic polarization and magnetization fields and the free charge and current densities through (\ref{eq:micro}), such a set $\{\boldsymbol{p}(\boldsymbol{x},t),\boldsymbol{m}(\boldsymbol{x},t),\rho_{F}(\boldsymbol{x},t),\bm{j}_{F}(\boldsymbol{x},t)\}$ satisfying (\ref{eq:micro}) is far from unique. Then it is not surprising that different approaches to defining such quantities can yield different results, and that the electronic contribution to the dipole moments (\ref{multipole}) describing the microscopic polarization (\ref{microP}) and magnetization (\ref{microM}) fields depend on the choice of Wannier frame (and therefore on the choice of ELWFs). In certain instances we have found \cite{Mahon2020,Mahon2020a} these quantities to be independent of this choice --- for example, the unperturbed $\bm{M}$ and the linearly induced $\bm{P}$ due to an electric field in a bulk trivial insulator initially occupying its zero-temperature ground state --- but these are to be considered special cases. In fact, in such a trivial insulator the unperturbed $\bm{P}$ and the linearly induced $\bm{P}$ ($\bm{M}$) due to a magnetic (electric) field display a similar ``gauge freedom," which gives rise to a ``quantum of indeterminacy'' in each of these cases; both of these quantities can be derived within the ``modern theories" \cite{Resta1994,Essin2010,Malashevich2010} and the approach implemented here \cite{Mahon2019,Mahon2020}, and agreement is found. In that setting we have also shown \cite{Mahon2020a} that the linearly induced charge and current densities due to electromagnetic fields that can vary both spatially and temporally are gauge invariant in that, when obtained from the appropriate combinations of induced multipole moments, the resultant expressions are independent of the choice of Wannier frame, even though the induced multipole moments of the (in this case nonuniform) macroscopic polarization and magnetization fields are generally not. We have also previously shown \cite{Mahon2019} that for a Chern insulator subject to a uniform dc electric field, the linearly induced $\bm{P}$ and $\bm{J}_{F}$ lead to the usual quantized anomalous Hall current via the macroscopic analog of (\ref{eq:micro}).
	
	In contrast, in the ``modern theories'' (and extensions thereof), notions of $\bm{P}$ and $\bm{M}$ are fundamentally macroscopic and aimed at the study of unperturbed insulators or those that are adiabatically perturbed by a uniform electromagnetic field. Moreover, an implicit principle of both modern theories is that, in finite-sized insulators, $\boldsymbol{P}$ and $\boldsymbol{M}$, when taken as the usual charge and current density dipole moments, are experimentally accessible; it is implicitly assumed that the bulk quantities should coincide with those \cite{Resta2005,Vanderbilt2009}. While this is a seemingly natural assumption, in the case of a Chern insulator it comes at the cost of an additional term in the bulk $\bm{M}$ that involves a chemical potential. As noted above, it seems that such an assumption must be examined in a case-by-case basis, since the relation between these quantities in bulk and finite-sized systems is not straightforward and considerations at the boundary are often important, even in trivial insulators. For example, the bulk topological magnetoelectric coefficient does not in general determine that of a thin film \cite{Qi2008,Wang2015}. 
	
	It is not then surprising that we find disagreement with the extension of the ``modern theory of magnetization,'' owing to the different underlying philosophies of these approaches. In fact, this is elucidated precisely by the additional terms that appear in the expressions for the unperturbed $\bm{M}$ of a bulk Chern insulator. In the ``modern theory," the additional term involves a chemical potential multiplied by a Berry curvature, which is best understood in the context of finite-size systems and is a consequence of the thermodynamic extension of $\bm{M}$ for a trivial insulator; in two dimensions, that term is related to the quantum anomalous Hall current by means of a Streda formula \cite{Resta2006}. In contrast, the $\bm{M}$ we derive for a bulk Chern insulator features an extra term involving the matrix $\mathcal{W}_{mn}^a(\bm{k})$ defined in (\ref{W}), which is an object directly related to the geometric structures encoding the ``gauge freedom" in our formalism. More precisely, associated with the Bloch bundle is its \textit{frame bundle} \cite{HamiltonBook}, the (local) sections of which are frames over open sets in the Brillouin zone; any two such frames are related pointwise by a unitary transformation. Examples include the locally defined Bloch frame $(\ket{n\bm{k}})_{n \in J}$ and the globally defined Wannier frame $(\ket{\alpha\bm{k}})_{\alpha \in J}$, and the unitary transformation relating them is precisely the transformation whose components are given in (\ref{pseudoBloch}). If we equip the frame bundle with a connection, which in turn induces a connection on the Bloch bundle, then these two frames yield (local) component representations of this connection; in the Bloch frame $(\ket{n\bm{k}})_{n \in J}$, the components of the Berry connection are given by $\xi_{mn}^a(\bm{k})$ defined in (\ref{eq:Bloch_connection}), while in the Wannier frame $(\ket{\alpha\bm{k}})_{\alpha \in J}$ its components are
	\begin{equation}
		\tilde{\xi}_{\beta\alpha}^a(\bm{k}) = \frac{i}{\Omega_{\text{uc}}} \int_{\Omega} d\bm{x}\, u_{\beta\bm{k}}^*(\bm{x})\frac{\partial u_{\alpha\bm{k}}(\bm{x})}{\partial k^a}.
		\label{connectionWannier}
	\end{equation}
	These two local representations (of the \textit{same} connection on the total space of the frame bundle) are related by
	\begin{widetext}
		\begin{align}
			\sum_{\alpha\beta}U_{m\beta}(\boldsymbol{k})\tilde{\xi}^a_{\beta\alpha}(\boldsymbol{k})U^{\dagger}_{\alpha n}(\boldsymbol{k})=\xi^a_{mn}(\boldsymbol{k})+\mathcal{W}^a_{mn}(\boldsymbol{k}).\label{connection}
		\end{align}
	\end{widetext}
	The dependence of these component representations of the Berry connection on the choice of frame is precisely encoded in $\mathcal{W}_{mn}^a(\bm{k})$, and so it seems natural that any expression in our formalism involving the Berry connection that differs from that of the ``modern theories,'' for example, the expression (\ref{nu}) for the orbital magnetization, will involve this matrix \footnote{From a rigorous perspective, the $\mathcal{W}_{mn}^a(\bm{k})$ are the components of the pullback of the Maurer-Cartan form on the structure group of the frame bundle (which is a principal bundle) to the base manifold $\text{BZ}$ by the (local) Bloch frame $(\ket{n\bm{k}})_{n\in J}$.}. In this sense the rich geometry behind topologically nontrivial phases like the Chern insulator plays a more prominent role in our formalism than it does in the ``modern theories."
	
	\section{Discussion}
	A unique feature of our microscopic formalism is the unified manner (\ref{eq:micro}) in which ``bound'' and ``free'' charge and current densities in crystalline solids are described. The latter are modelled via a generalized lattice gauge theory wherein site charges $Q_{\bm{R}}(t)$ are associated with the lattice sites $\bm{R}\in\Gamma$ and link currents $I(\bm{R},\bm{R}')$ between sites $\bm{R},\bm{R}'\in\Gamma$ are identified by evaluating the time evolution thereof \cite{Mahon2019}. In general, the microscopic free current density,
	\begin{equation}
		\bm{j}_F(\bm{x}) = \frac{1}{2} \sum_{\bm{R}\bm{R}'} \bm{s}(\bm{x};\bm{R}',\bm{R})I(\bm{R},\bm{R}'),
	\end{equation}
	arising from such link currents is nonvanishing for an unperturbed Chern insulator, which is evident upon examining the explicit form of $I(\boldsymbol{R},\boldsymbol{R}')$,
	\begin{widetext}
		\begin{equation}
			I(\boldsymbol{R},\boldsymbol{R}')=\frac{2e}{\hbar} \frac{\Omega_{\text{uc}}^2}{ (2\pi)^6}\sum_{nm\lambda\gamma}f_n\iint d\boldsymbol{k} d\boldsymbol{k}'\,E_{m\boldsymbol{k}'} \text{Im}\Big[e^{i(\boldsymbol{k}-\boldsymbol{k}')\cdot(\boldsymbol{R}'-\boldsymbol{R})} U_{n\lambda}(\boldsymbol{k})U^{\dagger}_{\lambda m}(\boldsymbol{k}')U_{m\gamma}(\boldsymbol{k}')U^{\dagger}_{\gamma n}(\boldsymbol{k})\Big].
			\label{linkcurrent}
		\end{equation}
	\end{widetext}
	This vanishes in the limit of a trivial insulator; choosing the unitary matrices $U_{n\alpha}(\bm{k})$ to be of block-diagonal form and introducing the filling factor $f_{\alpha}$ as in Sec.~\ref{Sec2A}, the result follows. The link currents (and thus the microscopic free current density $\boldsymbol{j}_{\text{F}}(\boldsymbol{x})$) are sensitive to the topology of the valence bundle for insulators -- they vanish for trivial insulators but are generically nonvanishing for Chern insulators. In contrast, $\boldsymbol{j}(\boldsymbol{x})$ is invariant under transformations of the form (\ref{pseudoBloch}), and therefore physically measurable (unlike $\boldsymbol{j}_{\text{F}}(\boldsymbol{x})$), and in general does not vanish for insulating and metallic systems. However, if a crystal possesses time-reversal symmetry, then $\boldsymbol{j}(\boldsymbol{x})$ vanishes; in this sense $\boldsymbol{j}(\boldsymbol{x})$ is explicitly sensitive to the symmetries rather than topology of the band structure.
	
	The macroscopic free current density, obtained from its microscopic analogue through spatial averaging, can be shown to vanish in the ground state, in agreement with assumptions of the ``modern theories.'' Even so, it does yield a nonvanishing contribution to the linear response of such Chern insulators in the presence of a uniform electric field \cite{Mahon2019} and thus plays a nontrivial role in phenomena such as the quantum anomalous Hall effect. Moreover, the free current density that we define is generally gauge dependent and therefore physically indeterminate in a bulk crystal. However, it may be the case that there is a correspondence between the role played by this bulk free current density and that of the so-called ``topologically protected'' surface states, which cross the Fermi energy in finite-sized Chern insulators \cite{VanderbiltChernInsulator}. In particular, the ``quantum of indeterminacy'' arising from the gauge dependence of the bulk quantities considered here is often argued to be related to the ambiguity of the surface configuration of a finite-sized sample \cite{VanderbiltBook}; for example, in the macroscopic polarization of a trivial insulator as well as the Chern-Simons contribution to the orbital magnetoelectric polarizability tensor. Moreover, it is known that the appearance of edge currents in finite systems is also closely related to the nonvanishing of the Hall conductivity tensor \cite{VanderbiltBook}. We therefore anticipate a physical connection between these edge currents and our bulk free current density when considering finite-sized systems. We intend on investigating this connection in a future publication.
	
	In summary, we have applied a previously developed microscopic formalism to a Chern insulator and derived formulas for the ground-state polarization (\ref{mu}) and orbital magnetization (\ref{nu}). Our expression for the polarization is formally similar to that of the ``modern theory,'' while our expression for the orbital magnetization is qualitatively different. In the ``modern theory'' the orbital magnetization of a Chern insulator is obtained by thermodynamic arguments that yield an explicit dependence on a chemical potential (the Fermi energy), even in the case of a bulk insulator. Meanwhile, our expression is derived from an underlying microscopic magnetization field defined in terms of an appropriate choice of ELWFs and therefore has manifest dependence on this choice.
	
	\section{Acknowledgments}
	
	We thank Rodrigo A.~Muniz for useful discussions. This work was supported by the Natural Sciences and Engineering Research Council of Canada (NSERC). P.~T.~M.~acknowledges an Ontario Graduate Scholarship.
	
	\appendix
	\section{Generalized site-quantity matrix elements}\label{appendixA}
	In the case of an unperturbed crystal considered here, the ``generalized site-quantity matrix elements" are obtained by first expanding the electron field operators in a basis of ELWFs:
	\begin{equation}
		\hat{\psi}(\bm{x},t) = \sum_{\alpha\bm{R}} W_{\alpha\bm{R}}(\bm{x}) \hat{a}_{\alpha\bm{R}}(t).
		\label{ELWFexpansion}
	\end{equation}
	\\
	The dynamics of the electron field operator $\hat{\psi}(\bm{x},t)$ are such that the differential operators associated with the spatial components of the conserved current take the usual form \footnote{See, \textit{e.g.}, Peskin and Schroeder \cite{Peskin}.}:
	\begin{align}
		J^a\big(\boldsymbol{x},\boldsymbol{\mathfrak{p}}(\boldsymbol{x})\big)=\frac{e}{m}\mathfrak{p}^a(\boldsymbol{x}),
	\end{align}
	\\
	where $\bm{\mathfrak{p}}(\bm{x})$ was defined in (\ref{frakturp}). In terms of the expansion (\ref{ELWFexpansion}), the generalized site-quantity matrix element $\rho_{\beta\bm{R}';\alpha\bm{R}''}(\bm{x},\bm{R})$ associated with the electronic charge density is
	\begin{widetext}
		\begin{equation}
			\rho_{\beta\bm{R}';\alpha\bm{R}''}(\bm{x},\bm{R}) = \frac{e}{2}\big(\delta_{\bm{R}\bm{R}'} + \delta_{\bm{R}\bm{R}''}\big) W_{\beta\bm{R}'}^*(\bm{x}) W_{\alpha\bm{R}''}(\bm{x}),
		\end{equation}
		while the generalized site-quantity matrix element $\bm{j}_{\beta\bm{R}';\alpha\bm{R}''}(\bm{x},\bm{R})$ associated with the current density is
		\begin{equation}
			\bm{j}_{\beta\bm{R}';\alpha\bm{R}''}(\bm{x},\bm{R}) = \frac{1}{2}\big(\delta_{\bm{R}\bm{R}'} + \delta_{\bm{R}\bm{R}''}\big) \Big[W_{\beta\bm{R}''}^*(\bm{x}) \Big(\bm{J}(\bm{x},\bm{\mathfrak{p}}(\bm{x}))W_{\alpha\bm{R}'}(\bm{x})\Big) + \Big(\bm{J}(\bm{x},\bm{\mathfrak{p}}(\bm{x}))W_{\beta\bm{R}''}(\bm{x})\Big)^* W_{\alpha\bm{R}'}(\bm{x})\Big].
		\end{equation}
	\end{widetext}
	
	To derive the generalized site-quantity matrix element associated with the itinerant current density, we previously defined the general site quantity
	\begin{equation*}
		K_{\bm{R}}(\bm{x},t) \equiv \frac{\partial \rho_{\bm{R}}^{\text{el}}(\bm{x},t)}{\partial t} + \bm{\nabla}\cdot\bm{j}_{\bm{R}}(\bm{x},t),
	\end{equation*}
	which is introduced in the presence of an applied electromagnetic field. This quantity is generally nonvanishing due to charges which can move between neighboring lattice sites, but the continuity equation is still satisfied by the \textit{total} charge and current densities, which leads to
	\begin{equation}
		\sum_{\bm{R}} K_{\bm{R}}(\bm{x},t) = 0.
		\label{sumK}
	\end{equation}
	
	For an unperturbed crystal occupying in its zero-temperature ground state considered here, this simplifies to 
	\begin{equation}
		K_{\bm{R}}(\bm{x}) = \bm{\nabla}\cdot\bm{j}_{\bm{R}}(\bm{x}),
	\end{equation}
	and the itinerant current density is then defined by 
	\begin{equation*}
		\tilde{\bm{j}}(\bm{x}) = - \sum_{\bm{R}} \int d\bm{y}\, \bm{s}(\bm{x};\bm{y},\bm{R}) K_{\bm{R}}(\bm{y}) - \bm{j}_F(\bm{x}).
	\end{equation*}
	Using the general identity (\ref{sumK}), it is easy to show that the total itinerant current density always shows no divergence,
	\begin{equation}
		\bm{\nabla}\cdot\tilde{\bm{j}}(\bm{x}) = 0,
	\end{equation}
	here independent of time. Moreover, we can write $\tilde{\bm{j}}(\bm{x})=\sum_{\bm{R}}\tilde{\bm{j}}_{\bm{R}}(\bm{x})$, where $\bm{\nabla}\cdot\tilde{\bm{j}}_{\bm{R}}(\bm{x}) = 0$. Using (\ref{itinerantcurrentsitequant}) applied to an unperturbed crystal, it can be shown that the corresponding generalized site-quantity matrix elements are of the form
	\begin{equation}
		\tilde{\bm{j}}_{\beta\bm{R}';\alpha\bm{R}''}(\bm{x},\bm{R}) = \frac{1}{2} \big(\delta_{\bm{R}\bm{R}'} + \delta_{\bm{R}\bm{R}''}\big) \tilde{\bm{j}}_{\beta\bm{R}';\alpha\bm{R}''}(\bm{x}),
	\end{equation}
	\\
	where the exact form of the expression for $\tilde{\bm{j}}_{\beta\bm{R}';\alpha\bm{R}''}(\bm{x})$ is complicated and is given in previous work \cite{Mahon2019}.
	
	\begin{widetext}
		\section{Unperturbed macroscopic polarization and magnetization}\label{appendixB}
		A useful identity \cite{Marzari2012} is
		\begin{align}
			\int d\bm{x}\, W^*_{\beta\boldsymbol{R}}(\boldsymbol{x})x^aW_{\alpha\boldsymbol{0}}(\boldsymbol{x})=\frac{\Omega_{\text{uc}}}{(2\pi)^{3}}\int_{\text{BZ}}d\boldsymbol{k}\, e^{i\boldsymbol{k}\boldsymbol{\cdot}\boldsymbol{R}}\tilde{\xi}^a_{\beta\alpha}(\boldsymbol{k}),
			\label{firstMoment}
		\end{align}
		where $\tilde{\xi}^a_{\beta\alpha}(\boldsymbol{k})$ was defined in (\ref{connectionWannier}). The components (\ref{connectionWannier}) of the non-Abelian Berry connection in the Wannier frame $(\ket{\alpha\bm{k}})_{\alpha \in J}$ are related to its components in the Bloch frame $(\ket{n\boldsymbol{k}})_{n \in J}$ through (\ref{connection}). Since all of our Wannier and Bloch frames are periodic over $\text{BZ}$, so too are all of the quantities appearing in (\ref{connection}). Another useful identity is
		\begin{align}
			\int d\bm{x}\,\psi^*_{n'\boldsymbol{k}'}(\boldsymbol{x})\mathfrak{p}^a(\boldsymbol{x})\psi_{n\boldsymbol{k}}(\boldsymbol{x})=\mathfrak{p}^a_{n'n}(\boldsymbol{k})\delta({\boldsymbol{k}-\boldsymbol{k}'}), \label{eq:full_integral}
		\end{align}
		with matrix elements found to be \cite{Mahon2020}
		\begin{align}
			&\mathfrak{p}^a_{n'n}(\boldsymbol{k})=\delta_{n'n}\frac{m}{\hbar}\frac{\partial E_{n\boldsymbol{k}}}{\partial k^{a}}+\frac{im}{\hbar}\big(E_{n'\boldsymbol{k}}-E_{n\boldsymbol{k}}\big)\xi^a_{n'n}(\boldsymbol{k}). \label{pMatrixElements}
		\end{align}
		
		Implementing previously derived expressions summarized in Appendix \ref{appendixA}, the electric dipole moment associated with a lattice site $\bm{R}$ of a Chern insulator is found to be
		\begin{align}
			\mu^i_{\boldsymbol{R}}&\equiv\int d\bm{x}\, p^i_{\boldsymbol{R}}(\boldsymbol{x}) \nonumber \\
			&=\sum_{\alpha\beta\boldsymbol{R}'\boldsymbol{R}''}\left(\int d\bm{y}\, (y^i-R^i)\rho_{\beta\boldsymbol{R}';\alpha\boldsymbol{R}''}(\boldsymbol{y},\boldsymbol{R})\right)\eta_{\alpha\beta}(\boldsymbol{R}'' - \boldsymbol{R}') \nonumber \\
			&=e\frac{\Omega_{\text{uc}}}{(2\pi)^3}\text{Re}\sum_{n}f_n\int_{\text{BZ}}d\boldsymbol{k}\, \sum_{\alpha\beta\boldsymbol{R'}}e^{i\boldsymbol{k}\cdot(\boldsymbol{R}'-\boldsymbol{R})}U_{n\beta}(\bm{k})\left(\int d\bm{y}\, W^*_{\beta\boldsymbol{0}}(\boldsymbol{y})y^iW_{\alpha\boldsymbol{R}'-\boldsymbol{R}}(\boldsymbol{y})\right)U^\dagger_{\alpha n}(\bm{k}) \nonumber \\
			&=e\frac{\Omega_{\text{uc}}}{(2\pi)^3}\text{Re}\sum_{n}f_n\int_{\text{BZ}}d\boldsymbol{k} \left(\xi^i_{nn}(\bm{k})+\mathcal{W}^i_{nn}(\bm{k})\right).
		\end{align}
		Meanwhile, the atomiclike contribution to the magnetic dipole moment associated with a lattice site $\bm{R}$ is
		\begin{align}
			\bar{\nu}^i_{\boldsymbol{R}}&\equiv\int d\bm{x}\, \bar{m}^i_{\boldsymbol{R}}(\boldsymbol{x}) \nonumber \\
			&=\frac{1}{2c}\epsilon^{iab}\sum_{\alpha\beta\boldsymbol{R}'\boldsymbol{R}''}\left(\int d\bm{y}\, (y^a-R^a)j^b_{\beta\boldsymbol{R}';\alpha\boldsymbol{R}''}(\boldsymbol{y},\boldsymbol{R})\right)\eta_{\alpha\beta}(\boldsymbol{R}''-\boldsymbol{R}') \nonumber \\
			&=\frac{e}{2mc}\frac{\Omega_{\text{uc}}}{(2\pi)^3}\epsilon^{iab}\text{Re}\sum_{n}f_n\int_{\text{BZ}}d\boldsymbol{k}\,\sum_{\alpha\beta\boldsymbol{R'}}e^{i\boldsymbol{k}\cdot(\boldsymbol{R}'-\boldsymbol{R})}U_{n\beta}(\boldsymbol{k})\left(\int d\bm{y}\, W^*_{\beta\boldsymbol{0}}(\boldsymbol{y})y^a\mathfrak{p}^b(\boldsymbol{y})W_{\alpha\boldsymbol{R}'-\boldsymbol{R}}(\boldsymbol{y})\right)U^\dagger_{\alpha n}(\boldsymbol{k}) \nonumber \\
			&=-\frac{e}{2\hbar c}\frac{\Omega_{\text{uc}}}{(2\pi)^3}\epsilon^{iab}\text{Im}\sum_{nm}f_n\int_{\text{BZ}}d\boldsymbol{k}\Big(E_{m\boldsymbol{k}}\xi^a_{nm}(\bm{k})\xi^b_{mn}(\bm{k})+\big(E_{m\boldsymbol{k}}-E_{n\boldsymbol{k}}\big)\mathcal{W}^a_{nm}(\bm{k})\xi^b_{mn}(\bm{k})-E_{n\boldsymbol{k}}\mathcal{W}^a_{nm}(\bm{k})\mathcal{W}^b_{mn}(\bm{k})\Big),
			%\label{nuAtomic}
		\end{align}
		while the itinerant contribution to the magnetic dipole moment associated with a lattice site $\bm{R}$ is 
		\begin{align}
			\tilde{\nu}^i_{\boldsymbol{R}}&\equiv\int d\boldsymbol{x}\, \tilde{m}^i_{\boldsymbol{R}}(\boldsymbol{x}) \nonumber \\
			&=\frac{1}{2c}\epsilon^{iab}\sum_{\alpha\beta\boldsymbol{R}'\boldsymbol{R}''}\left(\int d\boldsymbol{y}\, (y^a-R^a)\tilde{j}^b_{\beta\boldsymbol{R}';\alpha\boldsymbol{R}''}(\boldsymbol{y},\boldsymbol{R})\right)\eta_{\alpha\beta}(\boldsymbol{R}''-\boldsymbol{R}') \nonumber \\
			&=-\frac{e}{2\hbar c}\frac{\Omega_{\text{uc}}}{(2\pi)^3}\epsilon^{iab}\text{Im}\sum_{n}f_n\int_{\text{BZ}}d\boldsymbol{k}\, \Big(-iE_{n\boldsymbol{k}}\partial_a\xi^b_{nn}(\bm{k})+\sum_{m}\big(E_{n\boldsymbol{k}}-E_{m\boldsymbol{k}}\big)\mathcal{W}^a_{nm}(\bm{k})\xi^b_{mn}(\bm{k})-E_{m\boldsymbol{k}}\mathcal{W}^a_{nm}(\bm{k})\mathcal{W}^b_{mn}(\bm{k})\Big).
			%\label{nuItinerant}
		\end{align}
		These expressions can be reached equally well as a limit of those appearing in Mahon and Sipe \cite{Mahon2021}, assuming the existence of a band gap in the set of energy bands that are initially partially occupied there. Indeed, the proof presented in Appendix C provides justification for the integration by parts that must be performed in order to relate these results.
		
		\section{Shift of energy zero}\label{appendixC}
		Starting with (\ref{nu}), we find that under a shift of the energy zero $E_{n\bm{k}} \to E_{n\bm{k}} + \varepsilon$, the components of the macroscopic magnetization transform as
		\begin{equation}
			M^i \to M^i + \varepsilon \frac{e}{\hbar c} \sum_n f_n \int_{\text{BZ}} \frac{d\bm{k}}{(2\pi)^3}\epsilon^{iab}\Big(\partial_a \xi_{nn}^b(\bm{k}) + \partial_a \mathcal{W}_{nn}^b(\bm{k})\Big).
			\label{ZPE1}
		\end{equation}
	\end{widetext}
	The Brillouin zone is diffeomorphic as a manifold to the 3-torus $\mathbb{T}^3 = \mathbb{R}^3 / \mathbb{Z}^3$, equipped with the flat Euclidean metric inherited from $\mathbb{R}^3$ \cite{PanatiPisante}. The first term of the integrand in (\ref{ZPE1}) is the $i^{\text{th}}$ component of the curl of $\xi^a(\bm{k})$, which is the (matrix-valued) vector field on $\text{BZ}$ associated through the flat metric to the (Berry) connection $1$-form $\xi(\bm{k})$, with matrix components
	\begin{equation}
		\xi_{mn}(\bm{k}) = i \big(m\bm{k}\big|\partial_a n\bm{k}\big) \mathrm{dk}^a,
	\end{equation}
	where the components preceding the coordinate $1$-form $\mathrm{dk}^a$ were defined in (\ref{eq:Bloch_connection}). By interior multiplication with the volume form $\mathrm{dk}^1\wedge\mathrm{dk}^2\wedge\mathrm{dk}^3$, the curl of $\xi^a(\bm{k})$ can be related to the exterior derivative of $\xi(\bm{k})$ \cite{LeeBook}, namely,
	\begin{equation}
		\sum_{i=1}^3 \epsilon^{iab} \partial_a \xi^b(\bm{k})\, \iota_{\partial_i}\big(\mathrm{dk}^1\wedge\mathrm{dk}^2\wedge\mathrm{dk}^3\big) = \mathrm{d}\xi(\bm{k}),\label{ZPE3}
	\end{equation}
	where $\iota_{\partial_i}$ denotes interior multiplication with the coordinate vector field $\partial_i = \partial / \partial k^i$, and similarly,
	\begin{equation}
		\sum_{i=1}^3 \epsilon^{iab} \partial_a \mathcal{W}^b(\bm{k})\, \iota_{\partial_i}\big(\mathrm{dk}^1\wedge\mathrm{dk}^2\wedge\mathrm{dk}^3\big) = \mathrm{d}\mathcal{W}(\bm{k}).\label{ZPE4}
	\end{equation}
	
	The domain of integration in (\ref{ZPE1}) is $\text{BZ} \cong \mathbb{T}^3 \cong \mathbb{S}^1 \times \mathbb{S}^1 \times \mathbb{S}^1$. Under the identification $\mathbb{S}^1 \times \mathbb{S}^1 \times \mathbb{S}^1 \cong \mathbb{S}^1 \times \mathbb{T}^2$, define a triple of smooth embedding maps $\phi_{s}^{(i)}: \mathbb{T}^{2} \hookrightarrow \mathbb{T}^3$ for fixed $s \in \mathbb{S}^1$ ($1 \leq i \leq 3$) by \cite{Kaufmann}
	\begin{align}
		\phi_{s}^{(1)}(k_2, k_3) = (s, k_2, k_3),\nonumber \\
		\phi_{s}^{(2)}(k_1, k_3) = (k_1, s, k_3),\nonumber \\
		\phi_{s}^{(3)}(k_1, k_2) = (k_1, k_2, s).
	\end{align}
	The image of $\mathbb{T}^2$ under $\phi_{s}^{(i)}$ is a $2$-cycle $\mathbb{T}_{s}^2 \equiv \phi_{s}^{(i)}(\mathbb{T}^2)$, and varying $s \in \mathbb{S}^1$ defines a smooth foliation of the Brillouin zone by $2$-tori. Then, using the embedding map $\phi_{s}^{(i)}$ we can pull back any $2$-form on $\text{BZ}$ to one on $\mathbb{T}^2$.
	
	Suppose that $M$ is an embedded submanifold of $\text{BZ}$ with the induced metric $h$, associated volume form $\text{dvol}_h$, and unit normal vector field $N$. Given a vector field $X$ on $\text{BZ}$, one can show that \cite{LeeBook}
	\begin{equation}
		\big(\phi_M\big)^*\Big[\iota_X(\mathrm{dk}^1\wedge\mathrm{dk}^2\wedge\mathrm{dk}^3)\Big] = \eval{\delta(X, N)}_{M}\, \text{dvol}_h,
	\end{equation}
	where $\phi_M : M \hookrightarrow \text{BZ}$ is the embedding map. In particular, taking $M = \mathbb{T}_{s}^2$ with unit normal vector field $\partial_i$, we can replace the term in brackets with either of (\ref{ZPE3},\ref{ZPE4}) and, noting that $\delta(X,\partial_i)$ just picks out the $i^{\text{th}}$ component function of $X$, we have
	\begin{align}
		\big(\phi_{s}^{(1)}\big)^*\big(\mathrm{d}\xi(\bm{k})\big) &= \epsilon^{1ab} \partial_a \xi^b(\bm{k})\, \mathrm{dk}^2 \wedge \mathrm{dk}^3,\nonumber \\
		\big(\phi_{s}^{(2)}\big)^*\big(\mathrm{d}\xi(\bm{k})\big) &= \epsilon^{2ab} \partial_a \xi^b(\bm{k})\, \mathrm{dk}^3 \wedge \mathrm{dk}^1,\nonumber \\
		\big(\phi_{s}^{(3)}\big)^*\big(\mathrm{d}\xi(\bm{k})\big) &= \epsilon^{3ab} \partial_a \xi^b(\bm{k})\, \mathrm{dk}^1 \wedge \mathrm{dk}^2.
	\end{align}
	If we take the wedge product of the $i^{\text{th}}$ expression with the $1$-form $\mathrm{ds}$, then we obtain a $3$-form on $\text{BZ}$ that corresponds (up to a possible sign) to the measure $d\bm{k}$ in the notation of the main text, and so we can write (\ref{ZPE1}) as
	\begin{widetext}
		\begin{equation}
			M^i \to M^i + \varepsilon \frac{e}{8\pi^3 \hbar c} \sum\limits_{n} f_n \int_{\mathbb{S}^1} \mathrm{ds} \int_{\mathbb{T}^2}\big(\phi_{s}^{(i)}\big)^*\Big(\mathrm{d}\xi_{nn}(\bm{k}) + \mathrm{d}\mathcal{W}_{nn}(\bm{k})\Big).
			\label{temp9}
		\end{equation}
	\end{widetext}
	
	From this geometric perspective, the existence of ELWFs is equivalent to the Bloch bundle being topologically trivial as a \textit{holomorphic} Hilbert bundle \cite{Panati2017}. This means that we replace the $\text{BZ}$ by an appropriate \textit{complexification} $\text{BZ}_{\mathbb{C}}$, which we take to be the following complex ``strip''
	\begin{equation}
		\text{BZ}_{\mathbb{C}} \equiv \Big\{\bm{\kappa} \in \mathbb{C}^3 \; \; \Big| \; \; \text{Re}(\bm{\kappa}) = \bm{k} \in \text{BZ} \;,\; \abs{\text{Im}(\kappa_i)}< a\Big\}
	\end{equation}
	for some $a > 0$, which is related to the exponential rate of decay of the ELWFs. Since the Bloch bundle is trivial as a holomorphic Hilbert bundle, there exists a globally-defined Wannier frame $(\ket{\alpha\bm{\kappa}})_{\alpha \in J}$ that is itself holomorphic, it is this frame with which we construct the ELWFs \cite{PanatiPisante}, meaning that each state $\ket{\alpha\bm{\kappa}}$ (or associated quasi-Bloch function $u_{\alpha\bm{\kappa}}(\bm{x}) = \bra{\bm{x}}\ket{\alpha \bm{\kappa}}$) satisfies the Cauchy-Riemann equations. These are conveniently encoded in a \textit{Dolbeault operator} $\bar{\partial}$ \cite{Huybrechts}, whose action on a function $f : \text{BZ}_{\mathbb{C}} \to \mathbb{C}$ is
	\begin{equation}
		\bar{\partial} f = \sum_{i = 1}^{3} \mathrm{d}\bar{\kappa}^i \frac{\partial f}{\partial \bar{\kappa}^i},\label{dolbeault}
	\end{equation}
	where $\bar{\kappa}^i$ is the complex conjugate of $\kappa^i$. If $f$ is holomorphic, then the Cauchy-Riemann equations are equivalent to the condition $\bar{\partial} f(\bm{\kappa}) = 0$. In particular, the Wannier frame $(\ket{\alpha\bm{\kappa}})_{\alpha \in J}$ being holomorphic means that $\bar{\partial} \ket{\alpha\bm{\kappa}} = 0$, or equivalently, that $\bar{\partial} u_{\alpha\bm{\kappa}}(\bm{x}) = 0$, for every $\alpha$ and all $\bm{\kappa} \in \text{BZ}_{\mathbb{C}}$ \cite{Brouder}.
	
	In an open neighborhood of a given point in $\text{BZ}_{\mathbb{C}}$, we can write the analytic continuation of (\ref{pseudoBloch}) as
	\begin{equation}
		u_{\alpha \bm{\kappa}}(\bm{x}) = \sum_{n} U_{n\alpha}(\bm{\kappa}) u_{n\bm{\kappa}}(\bm{x}),
	\end{equation}
	and with the \textit{local} condition $\bar{\partial} u_{\alpha\bm{\kappa}}(\bm{x}) = 0$ we have
	\begin{equation}
		\sum_{n} \big(\bar{\partial} U_{n\alpha}(\bm{\kappa})\big) u_{n\bm{\kappa}}(\bm{x}) = - \sum_{n} U_{n\alpha}(\bm{\kappa}) \big(\bar{\partial} u_{n\bm{\kappa}}(\bm{x})\big).
	\end{equation}
	Multiplying both sides by $\big(u_{m\bm{\kappa}}(\bm{x})\big)^*$ on the left and $U_{\alpha n}^{\dagger}(\bm{\kappa})$ on the right, integrating over the unit cell $\Omega$ and summing over $\alpha$, we find
	\begin{equation}
		\sum_{\alpha} \big(\bar{\partial}U_{m\alpha}(\bm{\kappa})\big) U_{\alpha n}^{\dagger}(\bm{\kappa}) = - \big(m\bm{\bar{\kappa}}\big|\bar{\partial}n\bm{\kappa}\big).
	\end{equation}
	We can rewrite this in terms of the exterior derivative $\mathrm{d} = \partial + \bar{\partial}$, where $\partial$ is the Dolbeault operator with $\bar{\bm{\kappa}}$ being replaced by $\bm{\kappa}$ in (\ref{dolbeault}) \cite{Huybrechts}. Taking the second exterior derivative of this expression and imposing the reality condition $\bm{\kappa} = \bar{\bm{\kappa}}$ to restrict to $\text{BZ}$, we end up with the local result
	\begin{equation}
		\sum_{n} f_n \Big(\mathrm{d}\xi_{nn}(\bm{k}) + \mathrm{d}\mathcal{W}_{nn}(\bm{k})\Big) = 0.\label{ZPEtemp}
	\end{equation}
	This holds only on sufficiently small open sets in $\text{BZ}$ for which (\ref{pseudoBloch}) is well defined. 
	
	Denote by $\mathscr{D}_{\mathcal{B}}$ the locus of degeneracies in $\text{BZ}$; that is, $\mathscr{D}_{\mathcal{B}}$ is the set of points $\bm{k}_0 \in \text{BZ}$ for which $E_n(\bm{k}_0) = E_m(\bm{k}_0)$ for some $n,m$. We will assume that $\mathscr{D}_{\mathcal{B}}$ consists of finitely many disconnected components (points and lines of degeneracy). Enclose $\mathscr{D}_{\mathcal{B}}$ within a closed submanifold $\mathcal{N}$ of $\text{BZ}$ that deformation retracts onto the closure of $\mathscr{D}_{\mathcal{B}}$. Then $\text{BZ} \setminus \mathcal{N}$ is an open submanifold of $\text{BZ}$ that has a vanishing intersection with $\mathscr{D}_{\mathcal{B}}$. Let $\mathbb{S}_0^1$ denote those $s \in \mathbb{S}^1$ for which $\mathbb{T}_s^2 \cap \mathcal{N} = \emptyset$. Then $\mathbb{S}^1 \setminus \mathbb{S}_0^1$ consists of finitely many disconnected closed intervals \cite{Kaufmann}
	\begin{equation}
		\mathbb{S}^1 \setminus \mathbb{S}_0^1 = \coprod_{\alpha = 1}^K \big[s_{\alpha}, s_{\alpha + 1}\big],
	\end{equation}
	where $s_{K + 1} = s_{1}$ by periodicity of $s$. For each $1\leq i \leq 3$, define the following function
	\begin{equation}
		\mathcal{Q}^{(i)}(s) \equiv \sum_{n} f_n \int_{\mathbb{T}^2} \big(\phi_{s}^{(i)}\big)^*\Big(\mathrm{d}\xi_{nn}(\bm{k}) +  \mathrm{d}\mathcal{W}_{nn}(\bm{k})\Big),
	\end{equation}
	where $s \in \mathbb{S}^1$. Consider the integral
	\begin{widetext}
		\begin{equation}
			\int_{\mathbb{S}^1} \mathrm{ds}\, \mathcal{Q}^{(i)}(s) = \int_{\mathbb{S}^1\setminus \mathbb{S}_0^1} \mathrm{ds}\, \mathcal{Q}^{(i)}(s) + \int_{\mathbb{S}_0^1} \mathrm{ds}\, \mathcal{Q}^{(i)}(s) = \int_{\mathbb{S}^1\setminus \mathbb{S}_0^1} \mathrm{ds}\, \mathcal{Q}^{(i)}(s),\label{ZPEtemp2}
		\end{equation}
	\end{widetext}
	where the second equality follows from the fact that $\text{BZ}\setminus\mathcal{N}$ contains no degeneracies and we can therefore find a sufficiently fine open cover for which (\ref{ZPEtemp}) holds on each open set thereof. Thus, we are left with only the second integral in (\ref{ZPEtemp2}). Noting that the $3$-form
	\begin{equation}
		\mathrm{ds} \wedge \big(\phi_{s}^{(i)}\big)^*\Big(\mathrm{d}\xi(\bm{k}) + \mathrm{d}\mathcal{W}(\bm{k}) \Big)
	\end{equation}
	is closed on each component of $\mathcal{N}$ and therefore locally exact by the Poincar\'{e} lemma \cite{LeeBook}, we can use the generalized Stokes theorem to write the integral on the right side of (\ref{ZPEtemp2}) as a sum over the boundary values \cite{Kaufmann},
	\begin{equation}
		\int_{\mathbb{S}^1\setminus\mathbb{S}_0^1} \mathrm{ds}\, \mathcal{Q}^{(i)}(s) = \sum_{\alpha =1}^{K} \Big(\mathcal{I}(s_{\alpha + 1}) - \mathcal{I}(s_{\alpha})\Big) = 0
	\end{equation}
	by periodicity of $\mathcal{Q}^{(i)}(s)$ and $s_{K+1} = s_1$, where $\mathcal{I}(s_{\alpha})$ denotes the integral on the left evaluated at the boundary value $s_{\alpha}$. Hence the integral
	\begin{equation}
		\sum\limits_{n} f_n \int_{\mathbb{S}^1} \mathrm{ds} \int_{\mathbb{T}^2}\big(\phi_{s}^{(i)}\big)^*\Big(\mathrm{d}\xi_{nn}(\bm{k}) + \mathrm{d}\mathcal{W}_{nn}(\bm{k})\Big) = 0,
	\end{equation}
	\\
	and the result follows.
	
	\bibliographystyle{apsrev4-1}
	\bibliography{ChernInsulator.bib}
	
\end{document}